\documentstyle{article}

\makeatletter
\@addtoreset{equation}{section}
\makeatother
\def\+{\;+\;}
\def\-{\;-\;}
\def\*{\,\cdot\,}

\def\n{\noindent}

\def\u{\acute u}
\def\w{\acute w}

\def\aph{\acute\phi}
\def\ds{\displaystyle}
\def\R{\mbox{\sf R}}
\def\U{\mbox{\sf U}}
\def\ou{\mbox{\bf u}}
\def\ow{\mbox{\bf w}}
\def\oR{\mbox{\bf R}}

\begin{document}

\begin{titlepage}

\title{$3D$ symplectic map}

\author{
S. M. Sergeev\\
Branch Institute for Nuclear Physics, Protvino 142284, Russia.\\
E-mail: sergeev\_ms@mx.ihep.su
}

\date{February, 1998}

\maketitle
\abstract{
Quantum $3D$ $R$ -- matrix in the classical (i. e. functional)
limit gives a symplectic map of dynamical variables.
The corresponding $3D$ evolution model is considered.
An auxiliary problem for it is a system of linear equations
playing the role of the monodromy matrix in $2D$ models.
A generating function for the integrals of motion is constructed
as a determinant of the auxiliary system.
}

\bigskip
\noindent
PACS: 05.50; 02.10; 02.20\\
{\em Keywords:} Discrete space -- time evolution models;
$2+1$ integrability; Tetrahedron equation
\end{titlepage}
\newpage

\section{\underline{Introduction}}

In this paper we formulate a classical discrete space -- time model
(DSTM)\footnote{Refs. \cite{hirota,nk-kdv,fv-hirota} are the rather
uncompleted list of the publications which are to be mentioned},
which is in some sense the classical counterpart of the
Zamolodchikov -- Bazhanov -- Baxter $3D$ lattice spin model
\cite{zam-solution,bb-first,mss-vertex}.
One way to obtain a classical model
associated with ZBB one is to consider the
path integration motivated quasi-classical limit
of ZBB, this was done in Ref. \cite{kms-classic}.

But here we choose another way: we consider a symplectic map
given by the functional limit of the operator $3D$ $\oR$ -- matrix
\cite{s-qd,ms-modified}. In this approach we elude the usage of
the Hirota -- Miwa equation, because of
this map is the {\em local} symplectic map for three
pairs of dynamical variables.
This map we use to construct $3D$ DSTM
with dynamical variables belonging to the vertices of the
kagome lattice.

An auxiliary linear problem for such maps in the operator as well as
in the functional approaches were found recently in \cite{electric}.
This problem is an over - defined set of linear equations with
compatibility conditions giving the map. Analogous system of linear
equations for whole two dimensional lattice plays the role of the
monodromy matrix in $2D$ models, and remarkably, the determinant
of this system corresponds to the trace of the monodromy matrix
and gives a generating function for integrals of motion.

In the main part of this paper (section 2: ``The model'')
we formulate the evolution model,
introduce linear variables
and show how the integrals of motion can be derived.
An interpretation of the results and some aspects of further
investigations are given in section 3: ``Discussion''.

Section 2 contains all the essential results of the paper,
but in such short form they look artificially.
All cumbersome details concerning the derivation of the model,
its geometrical interpretation and correct application
of the original linear problem from Ref. \cite{electric} are
given in Appendix.

\section{\underline{The model}}

\subsection{\underline{Algebraic definition of the evolution}}

Consider a ${\mbox{\cal Z}}^2$ span of a two dimensional plane generated
by two independent shifts $a$ and $b$ such that point $(n,m)$
of the span we present as $P\;=\; a^n\cdot b^m$. Assign to each point
$P$ three pairs of dynamical variables:
\begin{equation}
\ds
[\,u_1(P),w_1(P),\;\;u_2(P),w_2(P),\;\; u_3(P),w_3(P)\,]\;.
\end{equation}
For this discrete space system introduce a discrete time evolution
\begin{equation}\label{mmm}
\ds
[u_k(n,P),w_k(n,P)]\;\;\mapsto\;\;
[u_k(n+1,P),w_k(n+1,P)]\;,
\;\;\;k=1,2,3\;,
\end{equation}
defined by the following relations:
\begin{eqnarray}
\ds & \left\{
\begin{array}{ccc}
\ds w_1(n+1,P) & = & \ds
{w_1(n,P)\;w_2(n,P)\+ w_2(n,P)\;u_3(n,P)\+\kappa^2\;
u_3(n,P)\;w_3(n,P)\over w_3(n,P)}\;,\\
&&\\
\ds u_1(n+1,P) & = & \ds
{u_1(n,P)\;u_2(n,P)\;w_2(n,P)\over
u_1(n,P)\;w_2(n,P)\+ u_2(n,P)\;w_3(n,P)\+\kappa^2\;
u_1(n,P)\;w_3(n,P)}\;,
\end{array}\right.&\label{ev1}\\
&&\nonumber\\
&&\nonumber\\
\ds & \left\{
\begin{array}{lcc}
\ds w_2(n+1,a\cdot P) & = & \ds
{w_1(n,P)\;w_2(n,P)\;w_3(n,P)\over w_1(n,P)\;w_2(n,P)\+
w_2(n,P)\;u_3(n,P)\+ \kappa^2\;u_3(n,P)\;w_3(n,P)}\;,\\
&&\\
\ds u_2(n+1,a\cdot P) & = & \ds
{u_1(n,P)\;u_2(n,P)\;u_3(n,P)\over w_1(n,P)\;u_2(n,P)\+
u_2(n,P)\;u_3(n,P)\+ \kappa^2\;u_1(n,P)\;w_1(n,P)}\;,
\end{array}\right.&\label{ev2}\\
&&\nonumber\\
&&\nonumber\\
\ds & \left\{
\begin{array}{ccc}
\ds w_3(n+1,b\cdot P) & = & \ds
{u_2(n,P)\;w_2(n,P)\;w_3(n,P)\over u_1(n,P)\;w_2(n,P)\+
u_2(n,P)\;w_3(n,P)\+ \kappa^2\;u_1(n,P)\;w_3(n,P)}\;,\\
&&\\
\ds u_3(n+1,b\cdot P) & = & \ds
{w_1(n,P)\;u_2(n,P)\+ u_2(n,P)\;u_3(n,P)\+ \kappa^2\;
u_1(n,P)\;w_1(n,P)\over u_1(n,P)}\;.
\end{array}\right.&\label{ev3}
\end{eqnarray}
Here $\kappa^2$ is an arbitrary parameter common for all $P$.
The main property of this evolution is that it conserves
the Poisson brackets
\begin{equation}\label{poisson}
\ds
\{u_k(n,P)\;,\;w_m(n,Q)\}\;=\;u_k(n,P)\;w_m(n,Q)\;
\delta_{k,m}\;\delta_{P,Q}\;,
\end{equation}
i. e. map (\ref{mmm}) is the symplectic map.

\subsection{\underline{Auxiliary linear problem}}

Introduce a set of auxiliary variables $\psi^*_k(n,P)$ assigned to
the points of the $2D$ lattice. Let the evolution of $\psi^*_k(n,P)$
be
\begin{equation}\label{psi13}
\ds
\psi^*_1(n+1,P)\;=\;\psi^*_3(n,P)\;,\;\;\;\;
\psi^*_3(n+1,P)\;=\;\psi^*_2(n,a^{-1}b^{-1}\cdot P)\;,
\end{equation}
and
\begin{eqnarray}
\ds
\psi^*_2(n+1,P) & = &
\psi^*_1(n,P)\;\kappa\;({u_3(n,P)\over u_1(n,P)\,w_2(n,P)}\,-\,
{w_1(n,P)\over u_2(n,P)\,w_3(n,P)})\nonumber\\
\ds &&
-\,\psi^*_3(n,a\cdot P)\;{\kappa^2\over u_1(n+1,P)\,w_3(n,P)}\nonumber\\
\ds &&
+\,\psi^*_3(n,P)\;{u_3(n,P)\over u_2(n+1,a\cdot P)}\nonumber\\
\ds &&
+\,\psi^*_3(n,b\cdot P)\;{w_1(n+1,P)\over w_2(n,P)}\;.\label{psi2}
\end{eqnarray}
Thus $\psi^*_k(n,P)$ are the linear variables.
Define functions $J^*_k(n,P)$ as follows:
\begin{eqnarray}
\ds
&J^*_1(n,P) =
\psi^*_1(n,P)w_1(n,P)-\psi^*_2(n,P)u_1(n,P)+\psi^*_3(n,aP)\kappa
+\psi^*_3(n,bP)\kappa u_1(n,P)w_1(n,P),&
\nonumber\\
&&\nonumber\\
\ds
&J^*_2(n,P) =
\psi^*_1(n,P)\kappa + \psi^*_2(n,a^{-1}P)\kappa u_2(n,P)w_2(n,P)
+\psi^*_3(n,P)w_2(n,P)-\psi^*_3(n,bP)u_2(n,P),&
\nonumber\\
&&\nonumber\\
\ds
&J^*_3(n,P) =
-\psi^*_1(n,P)u_3(n,P) + \psi^*_2(n,b^{-1}P)w_3(n,P)
+\psi^*_3(n,aP)\kappa + \psi^*_3(n,P)\kappa u_3(n,P)w_3(n,P).&
\nonumber\\
&&\label{co-currents}
\end{eqnarray}
One can verify easily
\begin{eqnarray}
\ds
J^*_1(n+1,P) & = &
J^*_2(n,P)\;{u_1(n+1,P)w_1(n+1,P)\over u_2(n,P)w_2(n,P)}\,+\,
J^*_3(n,P)\;{\kappa\over w_3(n,P)}\;,\nonumber\\
&&\nonumber\\
\ds
J^*_2(n+1,a\cdot P) & = &
J^*_1(n,P)\;{u_2(n+1,a\cdot P)\over u_1(n,P)}\,+\,
J^*_3(n,P)\;{w_2(n+1,a\cdot P)\over w_3(n,P)}\;,\nonumber\\
&&\nonumber\\
\ds
J^*_3(n+1,b\cdot P) & = &
-\,J^*_1(n,P)\;{\kappa\over u_1(n,P)}\,+\,
J^*_2(n,P)\;
{u_3(n+1,b\cdot P)w_3(n+1,b\cdot P)\over u_2(n,P)w_2(n,P)}
\;,\nonumber\\
&&\label{j-star-map}
\end{eqnarray}
Regarding $\psi^*_k(n,P)$ and $J^*_k(n,P)$ as components of
row vectors $\Psi^*(n)$ and $\mbox{\bf J}^*(n)$,
we obtain three formal matrix equations: first, Eq. (\ref{co-currents})
we rewrite as
\begin{equation}
\ds
\mbox{\bf J}^*(n+1)\;=\;\Psi^*(n)\*\mbox{\bf D}(n)\;,
\end{equation}
second, Eqs. (\ref{psi13},\ref{psi2}) we rewrite as
\begin{equation}
\ds
\Psi^*(n+1)\;=\;\Psi^*(n)\*\mbox{\bf K}(n)\;,
\end{equation}
and finally, Eq. (\ref{j-star-map}) -- as
\begin{equation}
\ds
\mbox{\bf J}^*(n+1)\;=\;\mbox{\bf J}^*(n)\*\mbox{\bf M}(n)\;.
\end{equation}
Hence
\begin{equation}
\ds
\mbox{\bf J}^*(n+1)\;=\;
\Psi^*(n)\*\mbox{\bf K}(n)\*\mbox{\bf D}(n+1)\;=\;
\Psi^*(n)\*\mbox{\bf D}(n)\*\mbox{\bf M}(n)\;,
\end{equation}
so the main relation of this subsection arises:
\begin{equation}\label{main}
\ds
\mbox{\bf K}(n)\*\mbox{\bf D}(n+1)\;=\;
\mbox{\bf D}(n)\*\mbox{\bf M}(n)\;.
\end{equation}

Note, there is some ambiguity in the definition of $\mbox{\bf K}$ and
$\mbox{\bf M}$. Namely, for (\ref{psi13}) given
$\psi^*_2(n+1,P)$ can be obtained in general from
\begin{equation}
\ds
J^*_1(n+1,P)\;=\;
J^*_1(n,P)\;\alpha(n,P)\,+\,
J^*_2(n,P)\;\beta(n,P)\,+\,
J^*_3(n,P)\;\gamma(n,P)
\end{equation}
with arbitrary $\alpha,\beta,\gamma$. For $\mbox{\bf K}$ given --
$\mbox{\bf M}$ can be restored directly.

\subsection{\underline{Finite lattice and integrals of motion}}

Reduce now the infinite $\mbox{\cal Z}^2$ span of
the two dimensional plane to the finite $\mbox{\cal Z}_N^2$
span of a two dimensional torus. Namely, let
\begin{eqnarray}
\ds
& u_k(n,a^N\cdot P)\;=\;u_k(n,b^N\cdot P)\;=\;u_k(n,P)\;,&
\nonumber\\
\ds
& w_k(n,a^N\cdot P)\;=\;w_k(n,b^N\cdot P)\;=\;w_k(n,P)\;, &
\label{periodical}
\end{eqnarray}
$k=1,2,3$, for all $P$. Obviously, these conditions
are conserved by evolution (\ref{ev1}-\ref{ev3}).
Pure periodic boundary conditions are natural for the dynamical
variables, while for auxiliary $\psi^*_k(n,P)$ one can
impose {\em quasiperiodical} boundary conditions:
\begin{equation}\label{quasiperiodical}
\ds
\psi^*_k(n,a^N\cdot P)\;=\;\psi^*_k(n,P)\;x\;,\;\;\;
\psi^*_k(n,b^N\cdot P)\;=\;\psi^*_k(n,P)\;y\;.
\end{equation}
Variables $x,y$ play the r\^ole of the spectral parameters.
Now all matrices $\mbox{\bf K,M,D}$ are finite
($3N^2\times 3N^2$) dimensional matrix functions on $x,y$.

Take now the determinants of the left and right hand sides of
Eq. (\ref{main}). It is easy to see that
\begin{equation}
\ds
\mbox{det}\;\mbox{\bf K}(n,x,y)\;=\;
x^{-N}\,y^{-N}\;\mbox{det}\;\mbox{\bf K}(n,1,1)\;=\;
x^{-N}\,y^{-N}\; K(n)\;,
\end{equation}
and
\begin{equation}
\ds
\mbox{det}\;\mbox{\bf M}(n,x,y)\;=\;
x^{-N}\,y^{-N}\;\mbox{det}\;\mbox{\bf M}(n,1,1)\;=\;
x^{-N}\,y^{-N}\; M(n)\;.
\end{equation}
Denote further
\begin{equation}
\ds
D(n,x,y)\;=\;\mbox{det}\;\mbox{\bf D}(n,x,y)\;.
\end{equation}
Hence Eq. (\ref{main}) gives
\begin{equation}
\ds
D(n+1,x,y)\;=\;{M(n)\over K(n)}\;D(n,x,y)\;.
\end{equation}
Taking the series decomposition of $D$,
\begin{equation}
\ds
D(n,x,y)\;=\;\sum_{l,m}\;x^l\,y^m\;D_{l,m}(n)\;,
\end{equation}
we obtain the integrals of motion as the ratios of $D_{k,m}$:
\begin{equation}
\ds
I_{l,m}[u_k(n,P),w_k(n,P)]\;=\;{D_{l,m}(n)\over D_{N,-N}(n)}\;,
\end{equation}
where $I_{l,m}$ are functionals of $u_k(n,P),w_k(n,P)$ and
\begin{equation}
\ds
I_{l,m}[u_k(n,P),w_k(n,P)]\;=\;
I_{l,m}[u_k(n+1,P),w_k(n+1,P)]\;.
\end{equation}

\section{\underline{Discussion}}

So, we have described the completely integrable classical
discrete space -- time model. This model can be quantized
(see Appendix for the details), and general approach giving
relation (\ref{main}) has a quantum counterpart. This gives
a way to construct a $3D$ analogue of the Bethe ansatz.
Several papers concerning eigenvectors and eigenfunctions
for $3D$ transfer matrices and evolution operators are to be mentioned
here : \cite{korepanov-BA} and \cite{ks-boson}.

Mention also the evolution operator for the affine Toda field theory
(ATFT),
given in Ref. \cite{kr-ats}. First, the classical limit of ATFT
is closely connected with our evolution: the limit case when
$\kappa^2=0$ and
$\ds
{u_3(P)\,w_3(P)\,u_2(P)\over u_1(P)\,w_1(P)\,w_2(P)}\;
\mapsto\;0$ for all $P$, turning the frame of reference $n,a,b$
appropriately and hiding one space direction into the rank of
$A_N$\footnote{This operation in terms of classical variables is
defined badly, and in operator language gives some central elements.},
one can obtain a map corresponding to ATFT. Second, the approach
to $3D$ models as to $2D$ models with high rank symmetry group
implies the tedious technique of the nested Bethe ansatz
while we try to avoid this using ``relativistic'' approach.

\newpage

\appendix

\noindent
{\Large\bf Appendix}

\section{\underline{Definition of map $\R$}}

Consider the following map:
\begin{equation}\label{map}
\ds
[ u_1, w_1; u_2, w_2; u_3, w_3]
\;\;\stackrel{\ds\R}{\mapsto}\;\;
[\u_1,\w_1;\u_2,\w_2;\u_3,\w_3]\;,
\end{equation}
where $\u_k,\w_k$ are the following functions of $u_k,w_k$:
\begin{eqnarray}\label{themap}
\ds
\w_1\;=\;
{w_1\,w_2\,+\,w_2\,u_3\,+\,\kappa^2\,u_3\,w_3\over w_3}&,&
\u_1\;=\;
{u_1\,u_2\,w_2\over u_1\,w_2\,+\,u_2\,w_3\,+\,\kappa^2\,u_1\,w_3}\;,
\nonumber\\
\ds
\w_2\;=\;
{w_1\,w_2\,w_3\over w_1\,w_2\,+\,w_2\,u_3\,+\,\kappa^2\,u_3\,w_3}&,&
\u_2\;=\;
{u_1\,u_2\,u_3\over w_1\,u_2\,+\,u_2\,u_3\,+\,\kappa^2\,u_1\,w_1}\;,
\nonumber\\
\ds
\w_3\;=\;
{u_2\,w_2\,w_3\over u_1\,w_2\,+\,u_2\,w_3\,+\,\kappa^2\,u_1\,w_3}&,&
\u_3\;=\;
{w_1\,u_2\,+\,u_2\,u_3\,+\,\kappa^2\,u_1\,w_1\over u_1}\;.
\end{eqnarray}
Inverse action, $\R^{-1}$, is given by
\begin{eqnarray}\label{inversemap}
\ds
w_1\;=\;
{\w_1\,\w_2\,\u_3\over\w_1\,\w_3\,+\,\u_3\,\w_3\,+\,\kappa^2\,\w_1\,\w_2}
&,&
u_1\;=\;
{\u_1\,\w_2\,+\,\u_2\,\w_3\,+\,\kappa^2\,\u_2\,\w_2\over\w_3}\;,
\nonumber\\
\ds
w_2\;=\;
{\w_1\,\w_3\,+\,\u_3\,\w_3\,+\,\kappa^2\,\w_1\,\w_2\over\u_3}
&,&
u_2\;=\;
{\u_1\,\w_1\,+\,\u_1\,\u_3\,+\,\kappa^2\,\u_2\,\u_3\over\w_1}\;,
\nonumber\\
\ds
w_3\;=\;
{\u_1\,\w_2\,+\,\u_2\,\w_3\,+\,\kappa^2\,\u_2\,\w_2\over\u_1}
&,&
u_3\;=\;
{\w_1\,\u_2\,\u_3\over\u_1\,\w_1\,+\,\u_1\,\u_3\,+\,\kappa^2\,\u_2\,\u_3}
\;.
\end{eqnarray}
The reader can easily recognize $\R$ in our evolution formulae
(\ref{ev1}--\ref{ev3}) up to some re-enumeration of final states'
coordinates.
The map $\R$ can be extracted from the operator function
$\oR$ (see \cite{ms-modified}). In few words,
in the operator formulation $\oR$ is a function
of three independent commutative pairs
$\ou_k,\ow_k$, $k=1,2,3$, such that
\begin{equation}\label{qrelation}
\ds \ou_k\*\ow_k\;=\;q\;\ow_k\*\ou_k\;,
\end{equation}
and
\begin{equation}\label{operatormap}
\ds
\acute{\ou}_k\;=\;\oR\*\ou_k\*\oR^{-1}\;,\;\;\;
\acute{\ow}_k\;=\;\oR\*\ow_k\*\oR^{-1}\;.
\end{equation}
Map (\ref{themap}) appears in the limit $q\rightarrow 1$,
$q^{1/2}\rightarrow -1$, $\ou\rightarrow u$, $\ow\rightarrow w$.
As the consequence of (\ref{qrelation}) and (\ref{operatormap}),
$\R$ conserves the Poisson brackets (\ref{poisson}).

Note that the map $\R$ obey the Functional
Tetrahedron Equation \cite{ks-fun,electric,oneparam,kks-fte},
this fact provides the integrability.

Few words concerning the structure of $\R$. Introduce
useful variables $\overline u,\overline w,u,w,s$ as follows:
\begin{equation}\label{overl}
\ds
\overline w\;=\; w_1\; w_2\;,\;\;\;
\overline u\;=\; u_2\; u_3\;,\;\;\;
s\;{\overline u\over\overline w}\;=\;{u_1\over w_3}\;;
\end{equation}
\begin{equation}
\ds
u\;=\;{w_3\over w_2}\;,\;\;\;
w\;=\;{w_1\over u_3}\;,\;\;\;
s\;{u\over w}\;=\;{u_1\over u_2}\;.
\end{equation}
Then $\R$ conserves $\overline u,\overline w$ and $s$:
\begin{equation}
\acute{\overline u}\;=\;\overline u\;,\;\;\;
\acute{\overline w}\;=\;\overline w\;,\;\;\;
\acute s\;=\;s\;;
\end{equation}
and $\R$ acts on $u,w$ as follows:
\begin{equation}\label{uwmap}
\ds
\u \; = \; {1\over u}\;
{1\+ w\+ \kappa^2\,u \over s\+ w\+ s\,\kappa^2\,u}\;,
\;\;\;
\ds
\w \; = \; {s\over w}\;
{1\+ w\+ \kappa^2\,u \over 1\+ w\+ s\,\kappa^2\,u}\;.
\end{equation}
Define the function $H$:
\begin{equation}\label{ham}
\ds
H(u,w) \; = \;
w\+{s\over w}\+\kappa^2\;(1\+ w)\;
({1\over u}\+ u\;{s\over w})\;.
\end{equation}
It can be easily verified,
\begin{equation}
\ds
H(u,w)\;=\;H(\u,\w)\;.
\end{equation}
Note, $H(u,w)=h$ is a genus one elliptic curve.
Thus $H(u,w)$ can be considered as the Hamiltonian, corresponding to
$\R$, and the evolution
\begin{equation}
\ds
H(u,w)\;\tau{d\,u(\tau)\over d\,\tau}\;=\;\{\,H(u,w)\,,\,u\,\}
\;,\;\;\;
H(u,w)\;\tau{d\,w(\tau)\over d\,\tau}\;=\;\{\,H(u,w)\,,\,w\,\}
\end{equation}
gives map (\ref{uwmap}) as the first rational point.

\section{\underline{Evolution}}

Define now the geometrical evolution model with the help of the map
(\ref{map},\ref{themap}).

Consider the kagome lattice formed by $3N^2$ oriented lines
as it is shown in Fig. A.

\begin{center}

\setlength{\unitlength}{0.25mm} 
\thicklines
\begin{picture}(440,300)
\put(  80 ,  60 ){\vector(1,0){280}}
\put(   0 , 140 ){\vector(1,0){440}}
\put(  80 , 220 ){\vector(1,0){280}}
\put(   0 , 120 ){\vector(1,1){160}}
\put(  80 ,  40 ){\vector(1,1){240}}
\put( 240 ,  40 ){\vector(1,1){160}}
\put( 200 ,  40 ){\vector(-1,1){160}}
\put( 360 ,  40 ){\vector(-1,1){240}}
\put( 440 , 120 ){\vector(-1,1){160}}
\put( 217 , 150 ){$P$}
\put(  37 , 147 ){\small $\tau_2\tau_3^{-1}\,P$}
\put( 357 , 147 ){\small $\tau_2^{-1}\tau_3\,P$}
\put( 117 ,  67 ){\small $\tau_1\tau_3^{-1}\,P$}
\put( 117 , 227 ){\small $\tau_1^{-1}\tau_2\,P$}
\put( 277 ,  67 ){\small $\tau_1\tau_2^{-1}\,P$}
\put( 277 , 227 ){\small $\tau_1^{-1}\tau_3\,P$}
\put(120,0){Fig. A. A fragment of the kagome lattice.}
\end{picture}

\end{center}

\vspace{0.5cm}

\n
The lattice consists of two types of triangles and one type of hexagons.
Choose the triangles marked in Fig. A for the enumeration of the
lattice elements. A shift of the triangle $P$ to any other triangle
of the same type is s superposition of two independent shifts,
and because of the existence of three similar directions of the kagome
lattice it is convenient to use the homogeneous notation for a
multiplicative shift operator
\begin{equation}\label{shift}
\ds
\tau_1^\alpha\*\tau_2^\beta\*\tau_3^\gamma\;\;:\;\;
\alpha+\beta+\gamma\;=\;0\;,
\end{equation}
with the action of $\tau_1^{-1}\tau_2$, $\tau_1^{-1}\tau_3$ and
$\tau_2^{-1}\tau_3$ defined in Fig. A.

Consider now triangle $P$ in details. Enumerate the vertices
of $P$ by the numbers $1,2$ and $3$. Define the evolution $\U$
of the lattice as the mutual change of all the $P$ -- type
triangles in the spirit of the Yang -- Baxter equivalence as it is
shown in Fig. B.

\begin{center}

\setlength{\unitlength}{0.25mm} 
\thicklines
\begin{picture}(450,300)
\put(00,50){
\begin{picture}(200,200)
\put(  10 ,  70 ){\vector(1,0){180}}
\put(  40 ,  10 ){\vector(1,2){90}}
\put( 160 ,  10 ){\vector(-1,2){90}}
\put(  70 ,  70 ){\circle*{10}}
\put( 130 ,  70 ){\circle*{10}}
\put( 100 , 130 ){\circle*{10}}
\put(  95 ,  85 ){$P$}
\put( 110 , 125 ){$1$}
\put( 135 ,  80 ){$2$}
\put(  60 ,  80 ){$3$}
\end{picture}}
\put(250,50){
\begin{picture}(200,200)
\put(  10 , 130 ){\vector(1,0){180}}
\put(  70 ,  10 ){\vector(1,2){90}}
\put( 130 ,  10 ){\vector(-1,2){90}}
\put(  70 , 130 ){\circle*{10}}
\put( 130 , 130 ){\circle*{10}}
\put( 100 ,  70 ){\circle*{10}}
\put( 110 ,  65 ){$1$}\put(75,140){$2$}\put(120,140){$3$}
\put(  85 ,  10 ){$\tau_1\,P$}
\put( 160 , 150 ){$\tau_3\,P$}
\put(  10 , 150 ){$\tau_2\,P$}
\end{picture}}
\put(210,140){\begin{picture}(30,50)
\put(10,10){$\mapsto$}
\put(15,30){$U$}
\end{picture}}
\put(150,0){Fig. B. The action of $U$.}
\end{picture}

\end{center}

\vspace{0.5cm}

\n
The lattice obtained is again the kagome lattice with the similar set
of $P$ -- type triangles marked as $\tau_k\,P$ in Fig. B.
We regard this change as the evolution from time $n$ to time $n+1$,
and obviously the time can be extracted from $P$ as a power of
$\tau_k$ (recall, translations (\ref{shift}) for a time fixed
are of zero power). Thus three shifts $\tau_1,\tau_2,\tau_3$
are the elementary shifts of a {\bf three dimensional} rectangular
lattice, and the evolution $\U$ is the saw transfer matrix
of the $3D$ lattice (see Refs. \cite{korepanov-BA,ks-boson}
for geometrical details).

Assign now the pair of dynamical variables $u_k(P),w_k(P)$
to vertex $k$ of triangle $P$.
Define the evolution for these variables
via the map $\R$ (\ref{map},\ref{themap}) as follows:
\begin{eqnarray}
\ds
\U & : & [u_1(P),w_1(P);u_2(P),w_2(P); u_3(P),w_3(P)]
\;\;\stackrel{\ds\R}{\mapsto}\nonumber\\
\ds &&
[u_1(\tau_1\,P_),w_1(\tau_1\,P);
 u_2(\tau_2\,P_),w_2(\tau_2\,P);
 u_3(\tau_3\,P_),w_3(\tau_3\,P)]\;.\label{umap}
\end{eqnarray}
Denote now for the shortness the translations for the time fixed as
\begin{equation}
\ds
a \;=\; \tau_1^{-1}\*\tau_2^{}\;,\;\;\;
b \;=\; \tau_1^{-1}\*\tau_3^{}\;,
\end{equation}
and exhibit the ``time'' as a power of $\tau_1$ in the decomposition
of three dimensional $P$ with respect to $\tau_1,a,b$:
\begin{equation}
\ds
P\;=\;\tau_1^n\cdot a^\alpha\cdot b^\beta\;\mapsto\;
(n,P=a^\alpha\cdot b^\beta)\;.
\end{equation}
with this definition of time and space coordinates,
map (\ref{umap}) gives exactly the evolution
(\ref{ev1}--\ref{ev3}).

\section{\underline{Basic auxiliary linear problem}}

A version of an auxiliary linear problem for $3D$ models was
suggested in Ref. \cite{electric}.
Here we recall it briefly for the case of commutative dynamical
variables and show how integrals of the motion can be derived
with a help of it.

\noindent
We start from the basic definition.

\noindent
Consider some vertex $V$ from our (kagome) lattice. Complete pair
$u,w$ assigned to $V$ by an ``internal current'' $\phi$,
while observable (or ``external'') currents we are intending to
measure in the faces. Internal current as produces
additive contributions into four external currents of four
faces surrounding the vertex.
The values of the contributions  are shown in Fig. C.

\begin{center}

\setlength{\unitlength}{0.25mm} 
\thicklines
\begin{picture}(450,300)
\put(125,70){
\begin{picture}(200,200)
\put(   0 ,   0 ){\vector( 1,1){200}}
\put( 200 ,   0 ){\vector(-1,1){200}}
\put( 100 , 100 ){\circle*{10}}
\put(  70 , 160 ){$-\,u\;\phi$}
\put(  90 ,  30 ){$w\;\phi$}
\put(  25 ,  95 ){$k\;\phi$}
\put( 135 ,  95 ){$k\;u\;w\;\phi$}
\end{picture}}
\put(150,0){Fig. C. The current vertex.}
\end{picture}

\end{center}

\vspace{0.5cm}

\n
The rules of the game are quite simple:
\begin{itemize}
\item any observable face current is the sum of the surrounding vertices
      contributions, and
\item the face current for any closed face is zero.
\end{itemize}
This indeed is the decent linear problem due to the following remarkable
feature: for a given geometry of a graph the relations between the outer
face currents (with the conservation laws being taken into account)
define the dynamical variables of the elements of the graph unambiguously.

Consider now the problem of the equivalence of two Yang -- Baxter --
type graph as in Fig. B. In Fig. D these graphs are redrawn
with the face currents' contributions marked out.

\begin{center}

\setlength{\unitlength}{0.25mm} 
\thicklines
\begin{picture}(450,300)
\put(00,50){
\begin{picture}(200,200)
\put(  10 ,  70 ){\vector(1,0){180}}
\put(  40 ,  10 ){\vector(1,2){90}}
\put( 160 ,  10 ){\vector(-1,2){90}}
\put(  70 ,  70 ){\circle*{10}}\put(50,80){$V_3$}
\put( 130 ,  70 ){\circle*{10}}\put(135,80){$V_2$}
\put( 100 , 130 ){\circle*{10}}\put(110,125){$V_1$}
\put(  95 ,  90 ){$\phi_h$}
\put(  95 , 190 ){$\phi_e$}
\put(  95 ,  20 ){$\phi_b$}
\put(  35 , 120 ){$\phi_c$}
\put( 155 , 120 ){$\phi_d$}
\put(   0 ,  20 ){$\phi_g$}
\put( 185 ,  20 ){$\phi_f$}
\end{picture}}
\put(250,50){
\begin{picture}(200,200)
\put(  10 , 130 ){\vector(1,0){180}}
\put(  70 ,  10 ){\vector(1,2){90}}
\put( 130 ,  10 ){\vector(-1,2){90}}
\put(  70 , 130 ){\circle*{10}}\put(35,140){$V'_2$}
\put( 130 , 130 ){\circle*{10}}\put(145,140){$V'_3$}
\put( 100 ,  70 ){\circle*{10}}\put(110,65){$V'_1$}
\put(  95 , 105 ){$\phi_a$}
\put(  95 , 185 ){$\phi_e$}
\put(  95 ,  20 ){$\phi_b$}
\put(   0 , 175 ){$\phi_c$}
\put( 185 , 175 ){$\phi_d$}
\put(  30 ,  60 ){$\phi_g$}
\put( 160 ,  60 ){$\phi_f$}
\end{picture}}
\put(210,140){\begin{picture}(30,50)
\put(10,10){$\sim $}
\end{picture}}
\put(150,0){Fig. D. The equivalence.}
\end{picture}

\end{center}

\vspace{0.5cm}

\n
First, the conservation laws for the closed faces $\phi_a=\phi_h=0$.
Second, the equivalence means that the outer currents
$\phi_b,...,\phi_g$ for the left ang right hand side graphs coincide.
Thus restoring the values of the vertex contributions into the
face currents, one obtains the following system:
\begin{eqnarray}
\ds \phi_h & := &
w_1\;\phi_1\+k\;\phi_2\-u_3\;\phi_3\;=\;0\;,
\nonumber\\
\ds \phi_a & := &
-\;\u_1\;\aph_1\+k\;\u_2\;\w_2\;\aph_2\+\w_3\;\aph_3\;=\;0\;,
\nonumber\\
&&\nonumber\\
\ds \phi_b & := &
\w_1\;\aph_1\;=\;w_2\;\phi_2\+k\;u_3\;w_3\;\phi_3\;,
\nonumber\\
\ds \phi_c & := &
k\;\aph_2\;=\;k\;\phi_1\+ k\;\phi_3\;,
\nonumber\\
\ds \phi_d & := &
-\;\u_3\;\aph_3\;=\;k\;u_1\;w_1\;\phi_1\- u_2\;\phi_2\;,
\nonumber\\
&&\nonumber\\
\ds \phi_e & := &
-\; u_1\;\phi_1\;=\;-\;\u_2\;\aph_2\+k\;\aph_3\;,
\nonumber\\
\ds \phi_f & := &
k\;u_2\; w_2\;\phi_2\;=\;k\;\u_1\;\w_1\;\aph_1\+ k\;\u_3\;\w_3\;\phi_3\;,
\nonumber\\
\ds \phi_g & := &
w_3\;\phi_3\;=\;k\;\aph_1\+ \w_2\;\aph_2\;.
\nonumber\\
\end{eqnarray}
This system is eight homogeneous linear equations for six
internal currents $\phi_k,\aph_k$. The initial graph
(let hand side of Fig. D) is described by two independent currents,
so the rank of $8\times 6$ matrix of the coefficients must be
$6\- 3\;=\;4$. This is the zero curvature condition, and the solution of
it is given by
\begin{eqnarray}
&\ds
\w_1\;=\; w_2^{}\; \Lambda_3^{}\;,\;\;\;
\u_1\;=\;\Lambda_2^{-1}\; w_3^{-1}\;,
&\nonumber\\
&\ds
\w_2\;=\; \Lambda_3^{-1}\; w_1^{}\;,\;\;\;
\u_2\;=\;\Lambda_1^{-1}\; u_3^{}\;,
&\nonumber\\
&\ds
\w_3\;=\; \Lambda_2^{-1}\; u_1^{-1}\;,\;\;\;
\u_3\;=\;u_2^{}\; \Lambda_1^{}\;,
&
\end{eqnarray}
with
\begin{eqnarray}
\ds
\Lambda_1 & = &
u_1^{-1}\; u_3^{}\+ u_1^{-1}\; w_1^{}\+\kappa^2\;w_1^{}\; u_2^{-1}\;,
\nonumber\\
&&\nonumber\\
\Lambda_2 & = &
u_1^{-1}\; w_2^{-1}\+ u_2^{-1}\* w_3^{-1}\+
\kappa^2\;u_2^{-1}\; w_2^{-1}\;,
\nonumber\\
&&\nonumber\\
\Lambda_3 & = &
w_1^{}\; w_3^{-1}\+ w_3^{-1}\; u_3^{}\+ \kappa^2\;w_2^{-1}\; u_3^{}\;.
\end{eqnarray}
This is exactly the map $\R$.

\section{\underline{Linear problem for the whole lattice}}

Consider now the current system for the whole kagome lattice.
All faces are closed, so all the face currents are zeros.

Describe the system of ``vacuum currents'' in details. As
it was mentioned, the lattice consists of two types of the triangles
and one type hexagons. The elementary ``puzzle'' is shown in Fig. E.

\begin{center}

\setlength{\unitlength}{0.25mm} 
\thicklines
\begin{picture}(400,300)
\put( 100 , 100 ){\line( 1, 1){150}}
\put( 300 , 100 ){\line(-1, 1){150}}
\put( 100 , 100 ){\line( 1,-1){ 50}}
\put( 300 , 100 ){\line(-1,-1){ 50}}
\put( 150 , 250 ){\line( 1, 0){100}}
\put( 150 , 150 ){\line( 1, 0){100}}
\put( 150 ,  50 ){\line( 1, 0){100}}
\put( 150 , 250 ){\circle*{10}}\put(100,260){$V_2(aP)$}
\put( 250 , 250 ){\circle*{10}}\put(255,260){$V_3(bP)$}
\put( 200 , 200 ){\circle*{10}}\put(210,195){$V_1(P)$}
\put( 150 , 150 ){\circle*{10}}\put(105,155){$V_3(P)$}
\put( 250 , 150 ){\circle*{10}}\put(255,155){$V_2(P)$}
\put( 100 , 100 ){\circle*{10}}\put( 30, 95){$V_1(b^{-1}P)$}
\put( 300 , 100 ){\circle*{10}}\put(310, 95){$V_1(a^{-1}P)$}
\put( 150 ,  50 ){\circle*{10}}\put( 85, 35){$V_2(b^{-1}P)$}
\put( 250 ,  50 ){\circle*{10}}\put(255, 35){$V_3(a^{-1}P)$}
\put( 180 ,  95 ){$J_3(P)$}
\put( 180 , 165 ){$J_1(P)$}
\put( 180 , 225 ){$J_2(P)$}
\put(110,0){Fig. E. An elementary puzzle.}
\end{picture}

\end{center}

\vspace{0.5cm}

\n
Thus for each ``puzzle'' three ``vacuum'' equations are:
\begin{eqnarray}
\ds
& J_1(P)\;\stackrel{def}{=} &
w_1(P)\,\phi_1(P)\+ \kappa\,\phi_2(P)\- u_3(P)\,\phi_3(P)\;=\;0
\\
&&\nonumber\\
\ds
& J_2(P)\;\stackrel{def}{=} &
 - u_1(P)\,\phi_1(P) + \kappa u_2(a P) w_2(a P) \,\phi_2(a P)
 + w_3(b P) \,\phi_3(b P)\;=\;0
\\
&&\nonumber\\
\ds
& J_3(P)\;\stackrel{def}{=} &
w_2(P) \,\phi_2(P) + \kappa u_3(P) w_3(P) \,\phi_3(P)
+ \kappa \,\phi_1(a^{-1} P) + \kappa \,\phi_3(a^{-1} P)
\nonumber\\
\ds &&  + \kappa u_1(b^{-1} P) w_1(b^{-1} P) \,\phi_1(b^{-1} P)
 - u_2(b^{-1} P) \,\phi_2(b^{-1} P)\;=\;0\;.\nonumber\\
&&\label{vacuum}
\end{eqnarray}
We are looking for a nontrivial solution of this system, i.e.
corresponding determinant must be zero. This means that
the following quadratic form is degenerative:
\begin{equation}\ds
M(\psi^*_k,\phi_k)\;=\;
\sum_{P,k}\;
\psi^*_k(P)\* J_k(P)\;\equiv\;
\sum_{P,k}\; J^*_k(P)\*\phi_k(P)\;.
\end{equation}
In this way we obtain the alternative equivalent system of $J^*_k(P)$
(\ref{co-currents}), used in sections 2.2 and 2.3,
with three types co-vectors $\psi^*_1(P),\psi^*_2(P),\psi^*_3(P)$.
The transition from the original linear problem $\phi\leftrightarrow J$
to the co-linear problem $\psi^*\leftrightarrow J^*$ is important,
because one can't define an evolution for $\phi$ (but hence
there is the ambiguity in the definition of the evolution
for $\psi^*$).

Introduce the translations for $\psi^*_k$:
\begin{equation}
\ds
T_a\*\psi^*_k(P)\;=\;\psi^*_k(a\,P)\;,\;\;\;
T_b\*\psi^*_k(P)\;=\;\psi^*_k(b\,P)\;.
\end{equation}
Then system (\ref{co-currents}) can be written as
\begin{eqnarray}
&\ds
J^*_1\;=\;
w_1\*\psi^*_1\-u_1\*\psi^*_2\+\kappa\;(T_a\+u_1w_1\;T_b)\*\psi^*_3
\;=\;0\;,&
\nonumber\\
&\ds
J^*_2\;=\;
\kappa\*\psi^*_1\+\kappa u_2w_2\;T_a^{-1}\*\psi^*_2\+(w_2\-u_2\;T_b)
\*\psi^*_3
\;=\;0\;,&
\nonumber\\
&\ds
J^*_3\;=\;
-u_3\*\psi^*_1\+w_3\;T_b^{-1}\*\psi^*_2\+\kappa\;(T_a\+u_3w_3)\*\psi^*_3
\;=\;0\;,&
\end{eqnarray}
thus the following $3\times 3$ block form of the ``monodromy
matrix'' $\mbox{\bf D}$ arises:
\begin{equation}
\ds
\mbox{\bf D}^*(n)\;=\;
\left(\begin{array}{ccc}
W_1(n) & -\,U_1(n) & \kappa\, T_a\,+\,\kappa\,U_1(n)\,W_1(n)\,T_b \\
&&\\
\kappa & \kappa\,U_2(n)\,W_2(n)\,T_a^{-1} & W_2(n)\,-\,U_2(n)\,T_b \\
&&\\
-\,U_3(n) & W_3(n)\,T_b^{-1} & \kappa\,T_a\,+\,\kappa\,U_3(n)\,W_3(n)
\end{array}\right)\;,
\end{equation}
where block notations
$U_k(n)\;=\;\mbox{diag}_P\; u_k(n,P)$,
$W_k(n)\;=\;\mbox{diag}_P\; w_k(n,P)$.

As an example one can consider the case $N=1$. The determinant
of $\mbox{\bf D}$ gives immediately all the invariants
(\ref{overl},\ref{ham}) of single $\R$.

\end{document}